# SURFACE CONDITIONING EFFECT ON VACUUM MICROELECTRONICS COMPONENTS FABRICATED BY DEEP REACTIVE ION ETCHING

*Alain Phommahaxay[1], Gaëlle Lissorgues[1], Lionel Rousseau[1], Tarik Bourouina[1], Pierre Nicole[2]*

[1] ESIEE Paris-ESYCOM, Noisy-le-Grand, France
[2] Thales Airborne Systems, Elancourt, France

**ABSTRACT**

Advances in material processing such as silicon micromachining are opening the way to vacuum microelectronics. Two-dimensional vacuum components can be fabricated using the microsystems processes. We developed such devices using a single metal layer and silicon micromachining by DRIE. The latter technological step has significant impact on the characteristics of the vacuum components. This paper presents a brief summary of electron emission possibilities and the design leading to the fabrication of a lateral field emission diode. First measurement results and the aging of the devices are also discussed.

## 1. INTRODUCTION

The development of silicon micromachining led to the development of a variety of MEMS devices. High Q resonators are for instance useful in selective filtering or sensor applications. However they need to be operated at low pressures to minimize viscous damping. On the other hand, vacuum electronics components can now be fabricated with similar processes. Associations of MEMS devices with vacuum electronics are thus possible and can enable new functions.

Using a single mask fabrication process, our work enables the rapid prototyping of vacuum micro-components. The first prototypes, including lateral diodes and triodes, have been produced with a low-cost aluminum process, consisting of one metal layer, one lithography step and silicon micromachining. The effects of the process optimization on the characteristics will be assessed. Early measurement results and first aging study will show some drift of their performances over time. New applications such as the real time monitoring of the vacuum level in various MEMS devices could use this effect. Different configurations using more suitable and stable field emitter materials such as tungsten will be studied in the future.

## 2. ELECTRON EMISSION AND VACUUM MICROELECTRONICS

Despite the progress made in solid-state electronics, vacuum tube-based components are still employed but only in very specific applications. Indeed some of their performances are still unattained by solid-state devices [1]. Systems sustaining high powers and frequencies still use vacuum tubes. Indeed, the scattering probability of electrons traveling in vacuum is very low, leading to a quasi ballistic transport. Devices with cut-off frequencies in the range of tens of gigahertz are feasible with such technologies. Moreover their radiation hardness and power handling make them indispensable in space and harsh environment applications.

### 2.1. Thermionic Emission

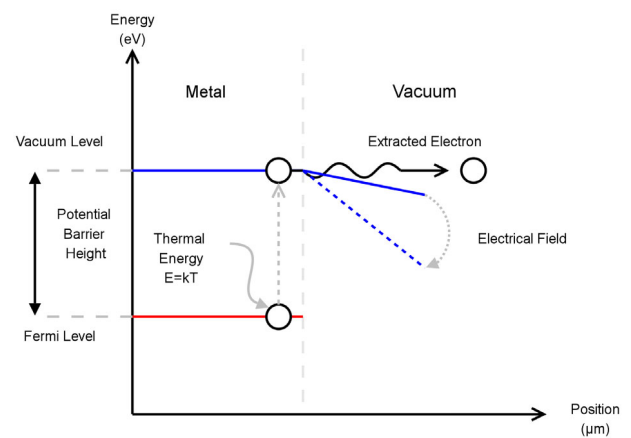

**Figure 1 : Thermionic Emission Principle**

Before the emergence of solid state electronics after the invention of the semiconductor transistors by Bardeen, Brattain and Shockley in the 1950s, most of the circuits were using vacuum tubes [4]. Since then, vacuum tubes disappeared from most applications mainly due to the difficulty to integrate them as their semiconductor counterparts. Indeed, vacuum tubes were using the thermo-ionic principle, shown in Figure 1, to generate a current flow. Electrons at the Fermi level have enough thermal energy to be extracted into vacuum. The corresponding current density is given by the Richardson-Dushmann equation [5]:

$$J = AT^2 e^{-\frac{\phi}{kT}}$$

where *A* is Richardson's constant, *T* is the temperature in Kelvin, $\phi$ is the metal work function and *k* is the Boltzmann constant.

The typical operating temperature of such tubes is thus above 800°C, making high integration density difficult. The advances in micro- and nano-fabrication techniques have led to a renewed interest in vacuum electronics. Indeed thanks to micromachining technology, their integration on silicon substrate is now possible, opening the way to vacuum microelectronics [1]. Micrometer and sub-micrometer-spaced electrodes can now be fabricated thanks to the dimensions achievable by lithography. High electric fields can be obtained by applying relatively small voltages, allowing electron tunneling and/or field emission. Such devices can furthermore be integrated with MEMS devices [2]-[3] allowing the design of relatively new functions.

### 2.1. Field Emission

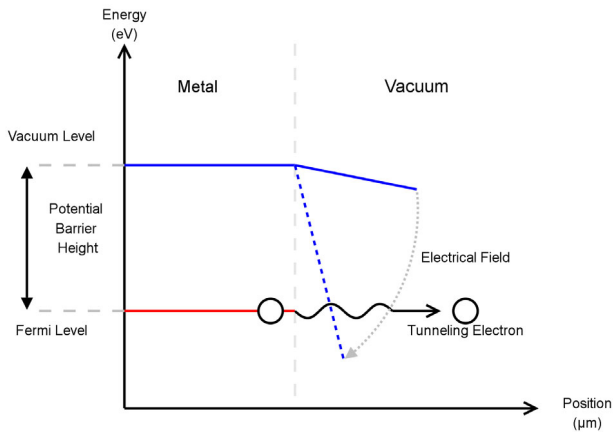

**Figure 2 : Field Emission Principle**

Contrary to thermionic emission, field emission does not require to heat a cathode. Indeed, it is based on the principle represented in Figure 2. Electrons tunnel from the Fermi level into vacuum under high electric field. Most part of the energy brought into the system is thus used to extract electrons, increasing the emission efficiency compared to the thermionic principle.

Field emission is principally governed by the Fowler-Nordheim theory, which uses the so-called triangular barrier approximation. The current density equation at 0°K is given by [1]:

$$J(F) \approx \frac{q}{4\pi^2 \hbar} \frac{\sqrt{\mu}}{[\mu+\phi]\sqrt{\phi}} F^2 \exp\left(-\frac{4}{3\hbar F}\sqrt{2m\phi^3}\right)$$

where *F* is the electric field, $\mu$ is the Fermi level, $\hbar$ is the Plank constant, *q* is the charge and *m* is the particle mass. In various works [6]-[7], this equation is simplified and leads to the following one:

$$J(F) \approx a_{FN} F^2 \exp(-b_{FN}/F)$$

where $a_{FN}$ and $b_{FN}$ are two constants depending on $\phi$ only if $\mu \approx \phi$, which is true for metals.

$$a_{FN} = \frac{K_1}{\phi}$$

, $K_1$ and $K_2$ constants.

$$b_{FN} = K_2 \phi^{3/2}$$

For a given electric field, the current density only depends on the material properties. The electrical field and the materials are the two parameters that can be tuned in order to increase the electron emission for a given geometry.

In fact $F=\beta V$, high electrical field values can be obtained by increasing either the voltage *V* or the field enhancement factor $\beta$, which is preferable for low voltage operation. This factor $\beta$ only depends on the geometry. The sharpest the tip is, the greater is this factor. The distance between electrodes can be reduced but it should be kept in mind that electrical breakdown can occur under intense fields [8].

New materials such as carbon nanotubes or diamond-like carbon either increase $\beta$ or decrease $\phi$. They allow the emission of electron with good uniformity and rather high current up to 1µA per tip. They are for example used in Field Emission Displays [9].

For a given geometry, the field enhancement factor can be tuned during the fabrication process. For a two-dimensional device fabricated on silicon, the Deep Reactive Ion Etching step would allow the tuning of this factor $\beta$, therefore changing the characteristics of the devices.

### 3. VACUUM MICRODIODES DESIGN

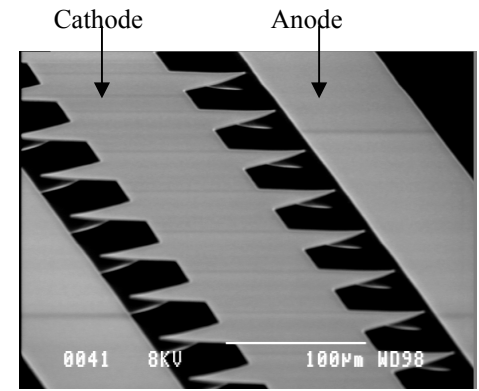

**Figure 3 : Vacuum Diodes Test Structure**

Early electrical measurements were planned to take place at atmospheric pressure, therefore low current levels were expected. Then, 20 parallel planar diodes as in Figure 3 were designed in order to increase the total emission current. The electrodes were spaced with 2µm-wide air-gap. The periodicity of the diode structure was chosen to minimize the self screening effect that can occur when two cathodes are too close to each other.

## 4. FABRICATION PROCESS

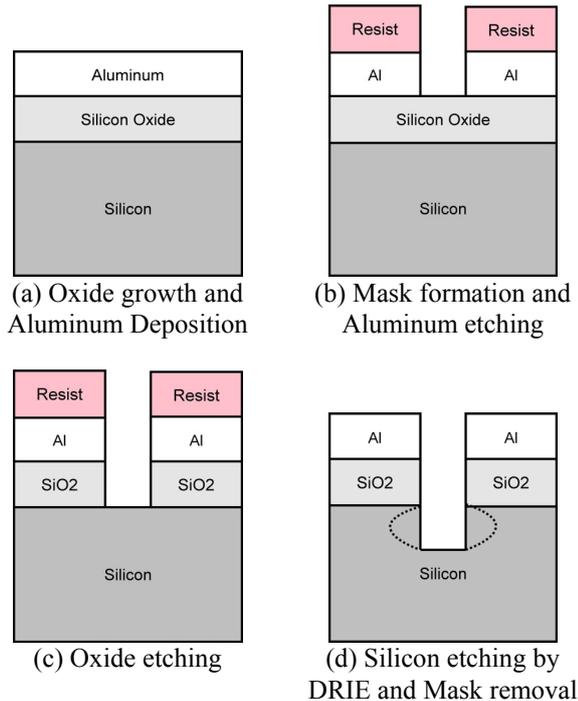

Figure 4 : Main Fabrication Process Steps

A schematic process flow is illustrated in Figure 4. Only one photomask combined with one metal layer are needed to achieve the fabrication of a lateral field emission diode. First of all, a thick thermal oxide was grown on the top of the silicon wafer. The early prototypes were fabricated with a 4000 Å $SiO_2$ layer on a P-type substrate. The metal was then deposited using either an electron beam evaporator or by sputtering as shown in Figure 4 (a). A 1 µm-thick aluminum film was used on our first devices. A hard mask was then formed by standard lithography and by etching the aluminum layer in chlorine plasma at step (b). The thermal oxide was then removed by fluorine anisotropic plasma etching through the mask as in Figure 4 (c). Deep Reactive Ion Etching, DRIE of silicon was then performed to release the emitter tip using an Alcatel 601E reactor. Varying process parameters allowed the control of the silicon etching profile. In fact, the length of the lateral etching and its deepness were easily controlled by alternating etching and passivation steps of the so-called Bosch process. The photoresist was finally stripped by O2 plasma at step (d).

As this structure tends to bend due to the residual stress of the silicon oxide layer, a misalignment between the two electrodes can happen. To avoid this, a High Resistivity Silicon (HRS) wafer can be used. Indeed the $SiO_2$ layer purpose is to ensure a good insulation between the metallic electrodes. This can also be done with HRS, and it simplifies the process since neither thermal oxidation nor oxide etching are needed.

This alternative process using HRS substrates can be very fast and offers prototypes in a matter of hours. The silicon oxide removal by plasma etching was indeed the slowest step of the process.

## 5. CHARACTERIZATIONS

### 5.1. Optical Characterization

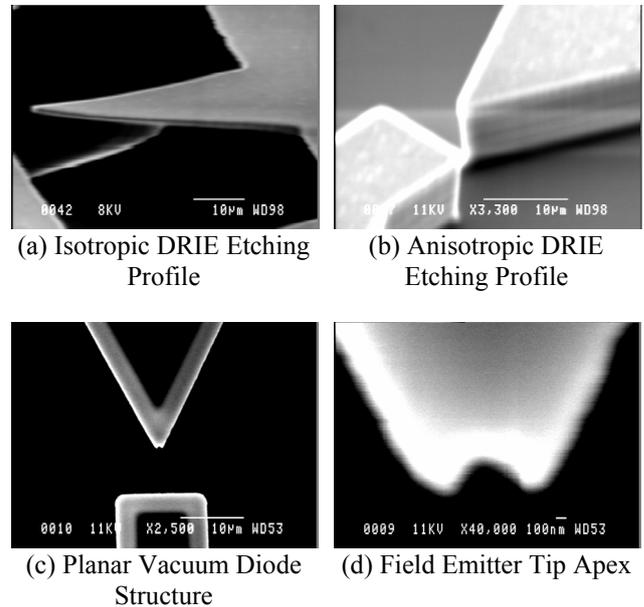

Figure 5 : SEM Picture of a Diode Structure Fabricated with Various DRIE Parameters

Optical characterizations were performed on the first prototypes. An interferometric optical profiler was used to control the silicon etching depth, while a scanning electron microscope was used to visualize the emitter profile and tip apex, as shown in Figure 5. Even though a standard lithography was used on the first samples, a 100nm-apex was found, which theoretically helps to enhance the electrical field.

### 5.2. Electrical Characterization

Preliminary measurements at atmospheric pressure were performed on the first prototypes. The on-wafer test setup illustrated in Figure 6 was composed of two electrical probes connected to the anode and the cathode. A voltage was applied through a Hewlett-Packard 4140B pico-amperemeter.

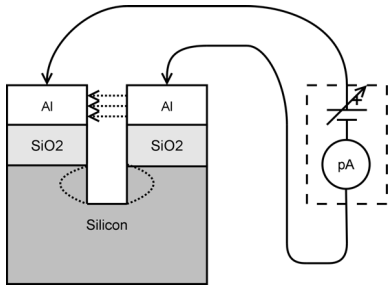

**Figure 6 : Electrical Test Setup**

### 5.3. Surface Conditioning Influence

Six months after their fabrication, some devices were characterized and then cleaned by acetone and orthophosphoric acid to remove the oxide layer formed on the aluminum surface. A very noisy response was recorded before the cleaning procedure. Current peeks during the measurement disappeared after the surface reconditioning as shown in Figure 7. However a smaller current value was detected, suggesting that impurities and protrusions can participate to the emission process.

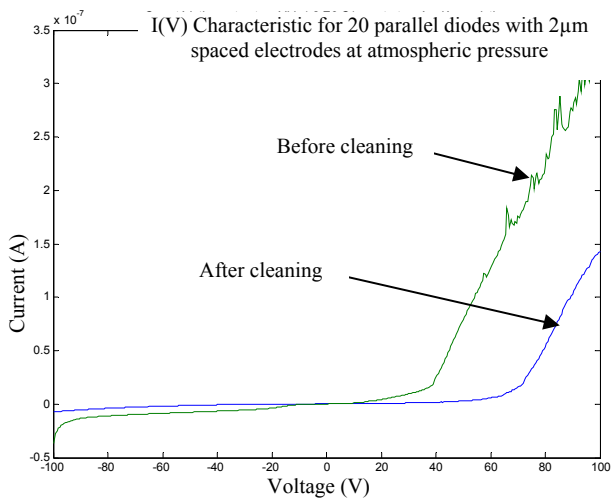

**Figure 7 : I(V) Curve of 20 Parallel Vacuum Diodes Before and After Surface Reconditioning**

The device composed of 20 parallel diodes with 2 micrometers-spaced electrodes showed a turn-on voltage of 25V and a current of 0.3µA when biased at 100V in atmospheric pressure before cleaning. The surface reconditioning increased the turn-on voltage to about 70V and reduced the current to 0.15µA at 100V. The devices are thus sensitive to the emitter surface chemistry and state.

Impurities deposited on non-encapsulated emitters can participate to the conduction, thus increasing the current but also instabilities given the high electric field between the anode and the cathode. Gases can also modify the emitter surface chemistry and the real work function. Oxygen found in the atmosphere can oxidize the surface of the conductor, consequently adding a barrier to the electron emission. Low work function materials are unfortunately more reactive chemically. Field emission devices must therefore be encapsulated in an inert environment such as nitrogen or even vacuum.

### 6. PACKAGING MONITORING

Pressures below $10^{-4}$ mbar are typically required to ensure good lifetimes and operating conditions in vacuum microelectronics [1]. The devices can still be operated at higher pressures but electrons would then collide with residual gases, possibly creating heavy ions, which would then bombard the emitters.

Such sensitivity to the pressure can allow the design of real-time monitoring structures inside micromachined vacuum cavities. The pressure level inside the cavity can be measured through a change of quality factor in MEMS resonating structures [10]-[11] or a frequency drift [12]. Field emitters would allow such monitoring without any mechanical structures. Indeed, the electron scattering probability and the surface chemistry are influenced by the gas type and pressure inside the cavity, thus modulating the current. New applications of our vacuum microelectronics devices could then be studied.

### 7. CONCLUSION

First prototypes of lateral vacuum microdiodes were fabricated and characterized at atmospheric pressure. The aging of the structures was studied. Changes in the current characteristics are mainly due to impurities and surface chemistry modifications. Such sensitivity to the environment can further be exploited in applications like real time vacuum monitoring in MEMS devices. Further experiments by encapsulating field emitters using either aluminum or other materials in an inert gas or in vacuum will be conducted in the near future. However this latter solution requires the use of getter materials, gas absorbers activated at high temperatures, to insure a low pressure level. Indeed, desorption will occur, various gases would be then trapped inside the cavity and increase the overall pressure.

## 8. AKNOWLEDGEMENT

This work is supported by the French Ministry of Defence (DGA) and Thales Airborne Systems. The authors would like to thank the ESIEE SMM team for their support and help during the fabrication process.